\documentclass[journal,twoside,web]{ieeecolor}
\usepackage{jsen}
\usepackage{amsmath,amssymb,amsfonts}
\usepackage{algorithmic}
\usepackage{graphicx}
\usepackage{textcomp}
\usepackage{wrapfig}
\usepackage{CJKutf8} %For Chinese Characters

\def\BibTeX{{\rm B\kern-.05em{\sc i\kern-.025em b}\kern-.08em
    T\kern-.1667em\lower.7ex\hbox{E}\kern-.125emX}}
\definecolor{abstractbg}{rgb}{0.89804,0.94510,0.83137}
\newcommand{\norm}[1]{\left\lVert#1\right\rVert}

 \DeclareMathOperator{\diag}{diag}

    \def \mS {\text{$\mathbf S$}}

\newcommand{\va}{\mathbf{a}}
\newcommand{\vg}{\mathbf{g}}
\newcommand{\vb}{\mathbf{b}}
\renewcommand{\mS}{\mathbf{S}}
\renewcommand{\norm}[1]{\left\lVert#1\right\rVert}

\newcommand{\fref}[1]{Fig. \ref{#1}}
\DeclareMathOperator{\lognorm}{lognorm}
\usepackage[pdftex]{hyperref}
\usepackage{subcaption}
\usepackage[
  style=ieee,
  dashed=false,
  hyperref=true,
  url=true,
  backend=biber,
  natbib=true
] {biblatex}
\author{Oliver Dürr, Po-Yu Fan, and Zong-Xian Yin
\thanks{Manuscript accepted for publication in IEEE Sensors, Mar 3, 2023;; 
For information on obtaining reprints of this article, please send  e-mail to: reprints@ieee.org. \ieeedoi}
\thanks{Oliver Dürr is with the Institute of Optical Systems at the University of Applied Sciences (HTWG) Konstanz (Germany)  (e-mail: oliver.duerr@htwg-konstanz.de).}
\thanks{Po-Yu Fan is with the Department of Computer Science and Information Engineering at the Southern Taiwan University of Science and Technology (STUST) (e-mail: mb1g0109@stust.edu.tw).}
\thanks{Zong-Xian Yin is the with Department of Computer Science and Information Engineering at the Southern Taiwan University of Science and Technology (STUST) (e-mail: yinzx@stust.edu.tw).}
\thanks{{\bfseries\textcopyright{}}
2023 IEEE. Personal use of this material is permitted. Permission from IEEE must be obtained for all
other uses, in any current or future media, including
reprinting/republishing this material for advertising or
promotional purposes, creating new collective works, for
resale or redistribution to servers or lists, or reuse of any
copyrighted component of this work in other works.}
}

\def\ieeedoi{Digital Object Identifier: \href{doi.org/10.1109/JSEN.2023.3272907}{10.1109/JSEN.2023.3272907}}

\title{Bayesian Calibration of MEMS Accelerometers}
\begin{document}
\markboth{\textcopyright 2023 IEEE. \ieeedoi}{}

\IEEEtitleabstractindextext{%
\fcolorbox{abstractbg}{abstractbg}{%
\begin{minipage}{\textwidth}%
\begin{wrapfigure}[12]{r}{1.5in}%
\includegraphics[width=1.49in]{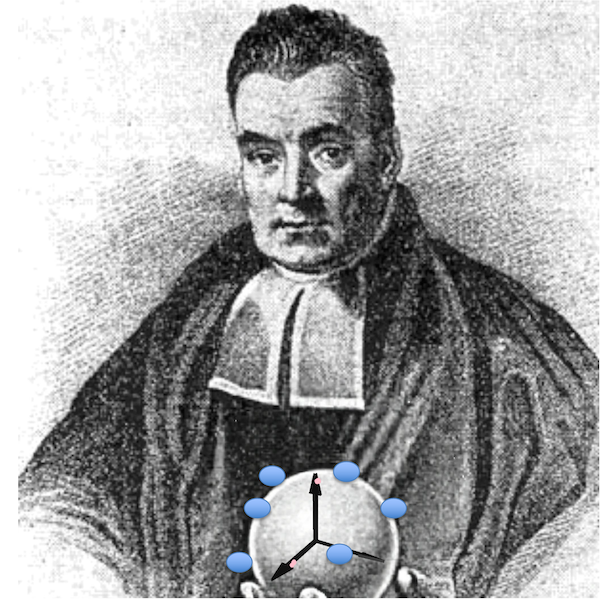}%
%%%%
%%Creation of figure. Using dal-e uploaded bayes from wikipedia. Asked del-e to hold a sphere and then copy pasted the calibration sphere from figure xxx and points using ppt 
\end{wrapfigure}%
\begin{abstract}
This study aims to investigate the utilization of Bayesian techniques for the calibration of micro-electro-mechanical systems (MEMS) accelerometers. These devices have garnered substantial interest in various practical applications and typically require calibration through error-correcting functions. The parameters of these error-correcting functions are determined during a calibration process. However, due to various sources of noise, these parameters cannot be determined with precision, making it desirable to incorporate uncertainty in the calibration models. Bayesian modeling offers a natural and complete way of reflecting uncertainty by treating the model parameters as variables rather than fixed values. Additionally, Bayesian modeling enables the incorporation of prior knowledge, making it an ideal choice for calibration. Nevertheless, it is infrequently used in sensor calibration. This study introduces Bayesian methods for the calibration of MEMS accelerometer data in a straightforward manner using recent advances in probabilistic programming.
\end{abstract}

\begin{IEEEkeywords}
%https://github.com/cgnorthcutt/ieee-keywords
Accelerometer calibration, 
gravity-based in-field calibration,
inertial measurement unit (IMU), 
low-cost IMU calibration,
micro-electro-mechanical systems (MEMS), 
multiposition calibration,
%%Check
Bayesian Parameter Estimation,
MCMC,
Uncertainty

%Enter key words or phrases in alphabetical 
%order, separated by commas. For a list of %suggested keywords, send a blank 
%e-mail to keywords@ieee.org or visit \underline
\end{IEEEkeywords}
\end{minipage}}}

\maketitle
\section{Introduction}
\IEEEPARstart{I}{n} 
\label{sec:introduction}
recent decades, micro-electro-mechanical systems (MEMS) accelerometers have gained significant attention in various practical applications, including monitoring physical activity \cite{boutenTriaxialAccelerometerPortable1997}, seismic activity detection \cite{jiangAllMetalOpticalFiber2013}, 
postural change \cite{wongMeasurementPosturalChange2009}, and gesture recognition \cite{benbasatInertialMeasurementFramework2002}. Especially consumer-level devices are prone to various sources of errors, such as misalignment and non-orthogonality of sensitivity axes, a drift of output signals over time, errors resulting from the aging of the silicone structure, and random noise \cite{ruMEMSInertialSensor2022}.

%Calibration in general
Calibration compares and corrects the output to match reference data over a range of output values and is an essential step in improving precision, especially for consumer sensors \cite{panahandehCalibrationAccelerometerTriad2010}. 
Traditionally, high-end inertial sensors have been calibrated through high-precision mechanical platforms. However, these non-autonomous methods are costly and require a controlled environment, which makes it challenging to use them for calibrating (MEMS) sensors. Therefore, researchers have explored alternative autonomous calibration methods that require less expensive equipment to make them useful for consumer-level devices. A recent review can be found in \cite{ruMEMSInertialSensor2022}. 

% All are Non Bayesian
Common to all methods (autonomous or not) is the estimation of the parameters of a sensor response function \cite{bergerBayesianSensorCalibration2022}. The response function can span a wide range of complexity, from simple linear transformations~\cite{panahandehCalibrationAccelerometerTriad2010} to complex neural networks \cite{liCalibNetCalibratingLowCost2021}. Various estimation and optimization techniques have been developed to determine these parameters, including least squares approaches and statistically sound and unbiased maximum likelihood methods \cite{panahandehCalibrationAccelerometerTriad2010, bergerBayesianSensorCalibration2022, ruMEMSInertialSensor2022}.

% Uncertainty
In the calibration process, the level of confidence in the predicted parameters should be considered. This uncertainty, known as aleatoric uncertainty, can be used, e.g., to determine whether sufficient data has been collected for the desired level of precision. Traditional statistical methods, such as confidence intervals, can be used in principle to quantify this aleatoric uncertainty. However, this is often neglected since it involves additional effort. Nevertheless, traditional statistical methods cannot account for the epistemic uncertainty arising when model assumptions are not met. Another drawback of traditional methods is that they cannot easily incorporate prior knowledge.

Bayesian methods offer a natural solution to these problems. Bayesian methods allow for the incorporation of prior knowledge and estimate the complete uncertainty (aleatoric and epistemic). These capabilities make Bayesian methods particularly well-suited for autonomous calibration. In the past, applying Bayesian methods required strong mathematical skills for constructing appropriate approximations. 
In recent years, however, there have been significant advancements in Bayesian methods, particularly with the availability of probabilistic programming frameworks like Stan~\cite{carpenterStanProbabilisticProgramming2017} and Pyro~\cite{binghamPyroDeepUniversal2019}, considerably easing the use of the Bayesian approaches.

Still, Bayesian methods for calibration are rarely employed in sensor calibration. We know only of two very recent publications, both using Bayesian inference for a Hall sensor system \cite{bergerBayesianSensorCalibration2022, bergerBayesianSensorCalibration2023}, significantly reducing the number of calibration points, and a single older reference cited in \cite{bergerBayesianSensorCalibration2022} in the context of experimental design \cite{craryBayesianOptimalDesign1995}. In contrast to \cite{bergerBayesianSensorCalibration2022} and  \cite{bergerBayesianSensorCalibration2023}, which both make distributional assumptions in the form of Gaussians, our study models the data generation process for MEMS accelerators directly; it makes no distributional assumptions but uses MCMC methods for inference instead.

%%%
The paper is organized as follows: In Section \ref{sec:error}, we motivate the error model assumed for MEMS accelerometers (\ref{sec:error-model}), discuss the calibration procedure (\ref{sec:cal}) used for sampling the calibration data, and briefly review traditional methods (\ref{sec-non-bayes}). Section \ref{sec-bayes}, summarizes the Bayesian methods used and introduces our full model and an approximation suitable for large amounts of calibration points. In (\ref{sec:res}), we discuss the results, first through simulations, to understand the properties of our method (\ref{sec:2D}) and verify the approximation and inference procedure (\ref{sec:3D}) before applying it to real data (\ref{sec:real}).

The code and data necessary to reproduce the results can be found at \url{https://github.com/oduerr/bayes_cal_paper}. 

\section{Error Model and calibration procedure}\label{sec:error}
\subsection{Error Model}\label{sec:error-model}
\subsubsection{Noiseless case}To understand the error model, we first ignore the stochastic noise. Several error models to calibrate the MEMS accelerometer sensors have been proposed in the past \cite{belkhoucheRobustCalibrationMEMS2022, boutenTriaxialAccelerometerPortable1997, hassanFieldCalibrationMethod2020, liCalibNetCalibratingLowCost2021, panahandehCalibrationAccelerometerTriad2010, qureshiAlgorithmInFieldCalibration2017}. All these linearly relate the three-dimensional output vector $\va$ of the measured acceleration to the true acceleration $\vg$ as follows
\begin{equation}
    \va = \mS \vg + \vb. \label{eq:error_det}
\end{equation}
The vector $\vb$ represents the bias, a fixed offset the sensor displays when no acceleration is detected. In case the matrix $\mS$ is diagonal, the components $\mS_{11}$, $\mS_{22}$, and $\mS_{33}$ are scale factors. Further, under some assumptions, it is possible to decompose $\mS$ into $\mS=\mathbf{T} \mathbf{K}$, where the diagonal matrix $\mathbf{K}$ contains the scale factors, and the upper-triangular matrix $\mathbf{T}$ contains three parameters corresponding 
 to misalignment, see \cite{qureshiAlgorithmInFieldCalibration2017} for further details. In practice, MEMS sensors are often calibrated with a diagonal matrix $\mS$ and the bias vector $\vb$, see e.g., \cite{belkhoucheRobustCalibrationMEMS2022}. While our method applies to the general case of arbitrary $\mS$, we focus here on the case of a diagonal $\mS$, leaving us with six interpretable parameters to be determined.
 \subsubsection{Adding noise}
In this work, we adopt a conventional approach by modeling the noise of the sensors as uncorrelated Gaussian noise. However, our approach is also suitable for modeling more complicated outcome distributions. I.e., we assume that the $i=1,2\ldots N$ observations of the accelerations $\va_i$ are independently drawn from a normal distribution $N$, centered around the mean value $\va$ from \eqref{eq:error_det}
\begin{equation}
    \va_i \sim N(\mS \vg_i + \vb,\mathbf{\Sigma}) \label{eq:error_noise}.
\end{equation}
Further, we assume, as usual, that the introduced noise in the different directions is independent and has the same magnitude, i.e., $\mathbf{\Sigma}=\sigma \mathbf{I}_3$, with $\mathbf{I}_3$ being the identity matrix. Under this assumptions and for diagonal $\mS=\tt{diag(S_1,S_2,S_3)}$ \eqref{eq:error_noise} can also be written as
\begin{equation}
    a_{ij} = S_j g_{ij} + b_j + \sigma \epsilon_{ij} \label{eq:error_noise2}\;,
\end{equation}
where $i=1,\ldots N$ enumerates the samples, $j=1,2,3$ the dimensions, and $\epsilon_{ij} \sim N(0,1)$ is independently drawn from a standard normal.

\subsection{Calibration Procedure}\label{sec:cal}
Most calibration procedures use the equivalence principle, which states that gravitational forces are indistinguishable from acceleration forces. Without any movement, sensors measure the current gravitation of the earth as acceleration. Since it is often impossible to determine the absolute value of the earth's gravity constant of the place where the calibration occurs, it is set to one. In this case, we measure the gravitation in units of the gravitational constant and $\norm{\vg} = 1$.

A particularly clever calibration procedure is the shape from motion method, which uses the fact that $\norm{\vg}$ is fixed (here we assume $\norm{\vg} = 1$) and does not require knowledge of the angles of the sensor w.r.t. earth's gravity \cite{kimCalibrationMultiAxisMEMS2007}. Recording the acceleration $\va_i$ at several sensor positions at rest is sufficient in these autonomous calibration methods. The precise angles are unnecessary; it is just necessary to ensure that the sensor does not move and, thus, no additional acceleration forces beyond gravity are present. The mathematical consequence of this calibration procedure is the constrain
\begin{equation}
     \norm{\vg_i} = 1 \label{eq:g1}\;.
\end{equation}

\subsection{Non-Bayesian approaches}\label{sec-non-bayes}
Several methods have been developed to estimate the parameters $\mS$, $\vb$, and $\mathbf{\Sigma}$ in \eqref{eq:error_noise} under constraint \eqref{eq:g1}. A straightforward approach is the maximum likelihood method. In general, the maximum likelihood method estimates the parameters $\theta$ of a model by choosing them so that the likelihood $p(D | \theta)$ of the observed data $D$ is maximized. 
\subsubsection{Full model}
In our case $\theta = \{\mS, \vb, \mathbf{\Sigma}\}$, the data $D=\{\va_i\}_{i=1,\ldots N}$ and the likelihood is given by
\begin{equation}
    p(D|\theta) = \prod_{i=1}^N N(\va_i; \mS \vg_i + \vb,\mathbf{\Sigma}) \label{eq:ml}\;.
\end{equation}
Due to the assumption of independence, the likelihood is given via the product of the densities $N(\va_i; \mS \vg_i + \vb,\mathbf{\Sigma})$ of the normal distribution at the observed accelerations $\va_i$. A direct optimization for calibration of \eqref{eq:ml} is done in \cite{sunCalibrationMEMSTriaxial2020}, where the constraint is modeled via Lagrange multipliers. The constraint \eqref{eq:g1} states that the coordinates of the gravitational acceleration must lie on a sphere. Introducing spherical coordinates, the radius $\norm{\vg}$, the polar angle $\vartheta \in [0,\pi]$ and the azimuthal angle $\varphi \in [0,2 \pi]$ allows to write $\vg$ as
\begin{equation}
    \vg = 
\begin{pmatrix}g_x \\ g_y \\ g_z \end{pmatrix} = \norm{\vg} \cdot 
\begin{pmatrix}
\sin(\vartheta) \cos(\varphi) \\ 
\sin(\vartheta) \sin(\varphi) \\ 
\cos(\vartheta)
\end{pmatrix}
\label{eq:sphere}
\end{equation}

In this coordinate system, the constrain $\norm{\vg_i}=1$ can be trivially fulfilled, and each data point adds only two new variables $\vartheta_i$ and $\varphi_i$. 
%\todo{Maybe: Few sensences about relation of ellipse fitting}
These direct approximations have many parameters: Beyond the parameters of interest, the unknowns $\vartheta_i$ and $\varphi_i$ also enter \eqref{eq:ml}. While it is possible to solve this equation, the many parameters might cause numerical problems, and suitable initial parameters must be chosen \cite{panahandehCalibrationAccelerometerTriad2010}.

\subsubsection{Radial approximation} The orthogonal distance regression (ODR) developed in \cite{splettEstimatingParametersCircles2019} reduces the number of parameters. In the spirit of standard regression (e.g. linear regression), one tries to solve \eqref{eq:error_det} w.r.t. $\vg$. Due to the noise present, we can only estimate $\vg$. For the ith sample, this estimate is
\begin{equation}
    \hat{\vg_i} = \mS^{-1}(\va_i - \vb)
\end{equation}
The radius is  $\norm{\hat{\vg}_i}=\sqrt{\hat{g}_{i1}^2 + \hat{g}_{i2}^2 + \hat{g}_{i3}^2}$, which is 1 for the noiseless case. For small values of $\sigma$ and small differences in the scale factors $S_j$, $\norm{\hat{\vg}_i}$ follows a Gaussian distribution $N(1,\sigma')$ centered around 1, which can be verified numerically. The standard deviation $\sigma'$ depends on the scales; for $S_1=S_2=S_3=1$ $\sigma' = \sigma$. %\todo{Here maybe find the correct distribution, in 2D it's Rice in 3D }. 
Given the measured values $\va_i$, the maximum likelihood solution maximizes the likelihood
\begin{equation}\label{eq:ml_ODR}
    L(\vb, \mS^{-1}, \sigma') = \prod_{i=1}^N N(1; \norm{\hat{g}_i}, \sigma') 
\end{equation}
by locating the optimal $\mS^{-1}$, $\vb$, and $\sigma'$. Since we assume a Gaussian, the maximum likelihood solution is equivalent to minimizing $\sum_i^N(\norm{\hat\vg_i} - \norm{\vg_i})^2$, i.e., the squared distances of the estimated radius $\norm{\hat\vg_i}$ from  $\norm{\vg_i}$, here taken to be 1  \cite{qureshiAlgorithmInFieldCalibration2017, wermanBayesianMethodFitting2001}. Note that the radial approximation is an approximation in two respects. First, the radius is non-Gaussian distributed, especially for larger values of $\sigma$. The second approximation is that, intuitively, the  maximum likelihood solution should find the minimal distance of a data point to the model. The proposed solution just minimizes the distance of $\hat{\vg}$ to the nearest point of the model and is, therefore is called orthogonal distance regression ODR \cite{splettEstimatingParametersCircles2019}. For a thorough examination of the approximation and a mathematically rigorous analysis of its constraints, see \cite{wermanBayesianMethodFitting2001}. 

\section{The suggested Bayesian approach}\label{sec-bayes}
In this section, we detail two Bayesian methods for calibration. They correspond to the "full model" and the introduced ODR approximation. We give a concise introduction to Bayesian statistics, as it is relevant for calibration (see e.g. \cite{mcelreath2020statistical} or \cite{gelmanBayesianDataAnalysis2015} for an introduction to modern Bayesian Statistics).
Generally, in the Bayesian approach, all parameters $\theta$ in a model are treated as random variables and thus have distributions. The distribution of a parameter reflects the degree of belief we have for particular values of this parameter. In the context of calibration, only continuous variables are of importance; therefore, we consider probability densities. Bayesian statistics starts with stating the prior belief. The first important probability density is the {\em prior} $p(\theta)$, which reflects our prior belief that a particular value $\theta$ of the parameter is taken. Seeing the data $D$, we update the prior distribution to the {\em posterior} $p(\theta|D)$, i.e., we condition on the data. The correct update rule is the Bayes formula
\begin{equation}
    p(\theta|D)  = \frac{p(D|\theta) p(\theta)}{\int p(D|\theta) p(\theta) \; d\theta} \label{eq:bayes}\;,
\end{equation}
which determines the posterior from the prior and the likelihood $p(D|\theta)$. The estimation of the optimal values of the parameters in mean squared error or maximum likelihood methods is now replaced by conditioning the prior on data. It is important to restate that the posterior is not a single value but a complete distribution that reflects our belief after seeing the data.

The integral in the denominator of \eqref{eq:bayes} can be calculated analytically only for elementary problems. This fact has hampered the broader use of Bayesian methods for decades. But thanks to the massive increase in computational power and further algorithmic developments, modern Bayesian approaches no longer need to rely on integration. The current gold standard is the Markov-Chain Monte-Carlo (MCMC) method. Instead of providing an analytical solution, MCMC methods produce samples $\theta_s$ from the posterior $p(\theta|D)$. Samples of advanced MCMC methods like Hamilton Monte Carlo (HMC)\cite{duaneHybridMonteCarlo1987} are nearly independent. Besides the likelihood and the prior itself, HMC methods also require their derivatives. In frameworks like Stan~\cite{carpenterStanProbabilisticProgramming2017} or Pyro~\cite{binghamPyroDeepUniversal2019}, the user just needs to define the likelihood, and the prior and all required calculations,  particularly the calculations of the derivatives, are done automatically. 
%For models with very many parameters, MCMC becomes less accessible, and variational inference methods or other approximations of the posterior have to be used.

Especially for many parameters, MCMC methods might run into numerical problems, which often can be solved by reformulating the problems or giving suitable initial values. To detect these problems, we examine trace plots \cite{gelmanBayesianDataAnalysis2015}, Gelman-Rubin statistic ($\hat{R}$) \cite{gelmanBayesianDataAnalysis2015}, and Effective Sample Size (ESS) statistic \cite{gelmanInferenceIterativeSimulation1992}. Trace plots reveal trends or patterns that indicate issues with convergence or mixing. The ESS statistic measures the number of independent samples that a chain generates. We consider chains to have converged if we find no apparent problems in the traceplots, the Gelman-Rubin statistic $\hat{R} < 1.10$ for all parameters, and an $\text{ESS} > 0.5 \cdot \text{number of samples}$. 

%depending on your needs a few 100 samples sufficient.  
\subsection{Bayesian model for the full model}\label{sec:full}
When setting up a Bayesian model, it is convenient to begin by describing the process by which the data is generated. For convenience, we introduce spherical coordinates. We start by defining the priors. 

\subsubsection{Priors} 
\begin{figure}[!t]
\centerline{\includegraphics[width=\columnwidth]{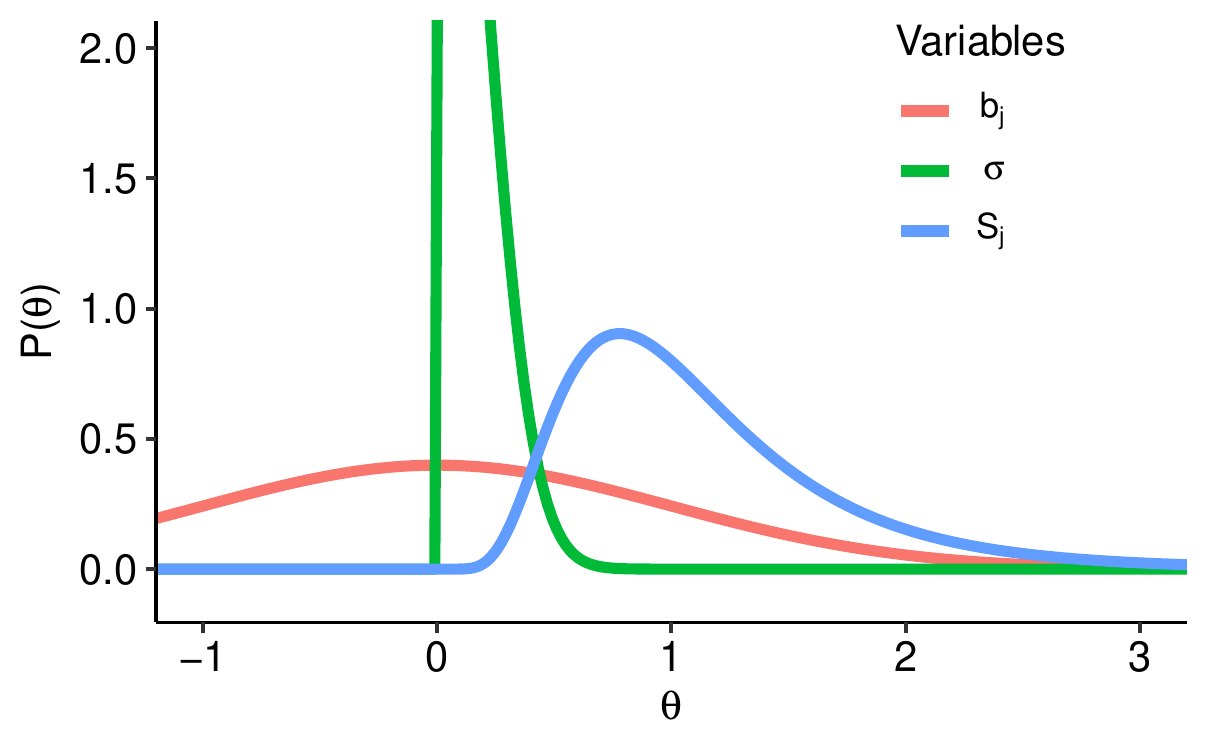}}
\caption{Priors used, shown are values for the bias terms $b_j \sim N(0,1)$, 
the noise term $\sigma \sim N(0, 0.2)$, and the scale $S_j \sim \lognorm(0,0.5)$. The scales are relative compared to the gravity $\norm{\vg}=1$}
\label{fig:prior}
\end{figure}

The sensor is at rest and experiences the acceleration $\vg$, of which we do not know the direction. Therefore, we use the uniform distribution $U$ for the prior, i.e., $p(\varphi)=U(0,2 \pi)$ and $p(\vartheta)=U(0, \pi)$ to describe our prior knowledge. 
The gravity constant is set to one (see \eqref{eq:g1}), which sets the scale for the other remaining variables ($\vb$, $\mS$, and $\sigma$). 
For these, we choose weakly informative priors with 
at least some mass around extreme but plausible values.\cite{gabryVisualizationBayesianWorkflow2019}.% 
We do not know the exact values of the bias term $\vb$, but we assume that the data is somewhat standardized around zero. A typical choice for such a variable is the standard normal. Since we do not believe that the different components of $b$ are correlated, we assume $b_j \sim N(0,1)$.
We further assume that $S_j$ follows a lognormal distribution with parameters $(0,0.5)$ (the typical effect is around one, and the scales are multiplicative). Finally, the noise level is small compared to the other magnitudes, so we set $\sigma$ in \eqref{eq:error_noise2} to 0.2. These prior distributions are displayed in \fref{fig:prior}. As the number of observations $N$ increases, the impact of the weakly informative priors diminishes, and the priors become less influential \cite{gabryVisualizationBayesianWorkflow2019}. These generic prior assumptions can be adapted if more information regarding the specific sensor is available, mainly when dealing with limited data. In such cases, the introduction of hyperpriors might also be beneficial. 
%For $N \rightarrow \infty$ the Bayesian approach generally converges to the maximum likelihood solution.  

\subsubsection{Likelihood}
To define the likelihood $p(\va_i | \mS, \vb, \sigma, \varphi_i, \vartheta_i)$, we need to determine the probability of an observation $\va_i$ given the parameters $\mS, \vb, \sigma, \varphi_i, \vartheta_i$. This can be seen as a generative process. First, starting from $\varphi_i, \vartheta_i$, we can calculate $\vg_i$ for the ith data point using \eqref{eq:sphere}. Then, given $\mS, \vb, \sigma$, we can calculate the likelihood $p(\va_i | \mS, \vb, \sigma, \varphi_i, \vartheta_i)$. This is reflected in the Stan-code in Appendix \ref{sec:stan-full}. 
%A sketch in plate notation for the model is given in \todo{DO}. 
A small particularity is that the model for the stationary accelerometer \eqref{eq:error_det} is invariant under rotation; hence, it is only defined up to a rotation. To make the model identifiable, we fix the angles $\vartheta_1$ and $\varphi_1$.

%%%%%%% Prior distributions %%%%%% 
%%%
% Figure is created with calibration_2d
%%%%%%%%%%%%%%%%%%%%%%%%%%%%%%%%%%%%%

Each new datapoint $i$ adds two variables $\vartheta_i$ and $\varphi_i$. For a diagonal matrix $\mS$, the model thus has $7+ 2\cdot (N-1)$ parameters, with $N$ being the number of data points. Although there is no principal limit in the number of data points, this model can be applied to, we experienced slow convergence of the Markov chains starting at around $N\approx 30$. To formulate a Bayesian approach applicable to large datasets, we use the ODR approximation.    
%%%%%%% Plate Notation %%%%%% 
%\begin{figure}[!t]
%\centerline{\includegraphics[width=\columnwidth]{figures/plateODR.pdf}}
%\label{fig:plate}
%\caption{Plate notation of the ODR approximation (left) and full model (right)  \todo{right create plate}). The observed quantities are shown in shaded circles, while open circles are unobserved parameters.}
%\end{figure}
%%%
% Figure is created with calibration_2d
%%%%%%%%%%%%%%%%%%%%%%%%%%%%%%%%%%%%%
\subsection{Bayesian model for ODR approximation}\label{sec:ODR} 
In this approximation, we consider the likelihood \eqref{eq:ml_ODR} and treat the parameters $\mS^{-1}, \vb$, and $\sigma'$ as Bayesian variables. We choose the same priors for $\mS, \vb, \sigma'$ as in \ref{sec:full}. This approach has only seven parameters and does not depend on the number of data points (Stan code is given in Appendix \ref{sec:stan-ODR}). It is important to note that the default initialization of Stan leads to non-mixing chains for many datapoints $N$, starting at $N \approx 100$. We, therefore, set the initial values for the MCMC simulation to $S_j=1$, $b_j = 0$ for $j=1,2,3$, and $\sigma' = 0.01$. 
\section{Results}\label{sec:res}
\subsection{2D Simulations Results}\label{sec:2D}
\begin{figure}[!t]
\centerline{\includegraphics[width=\columnwidth]{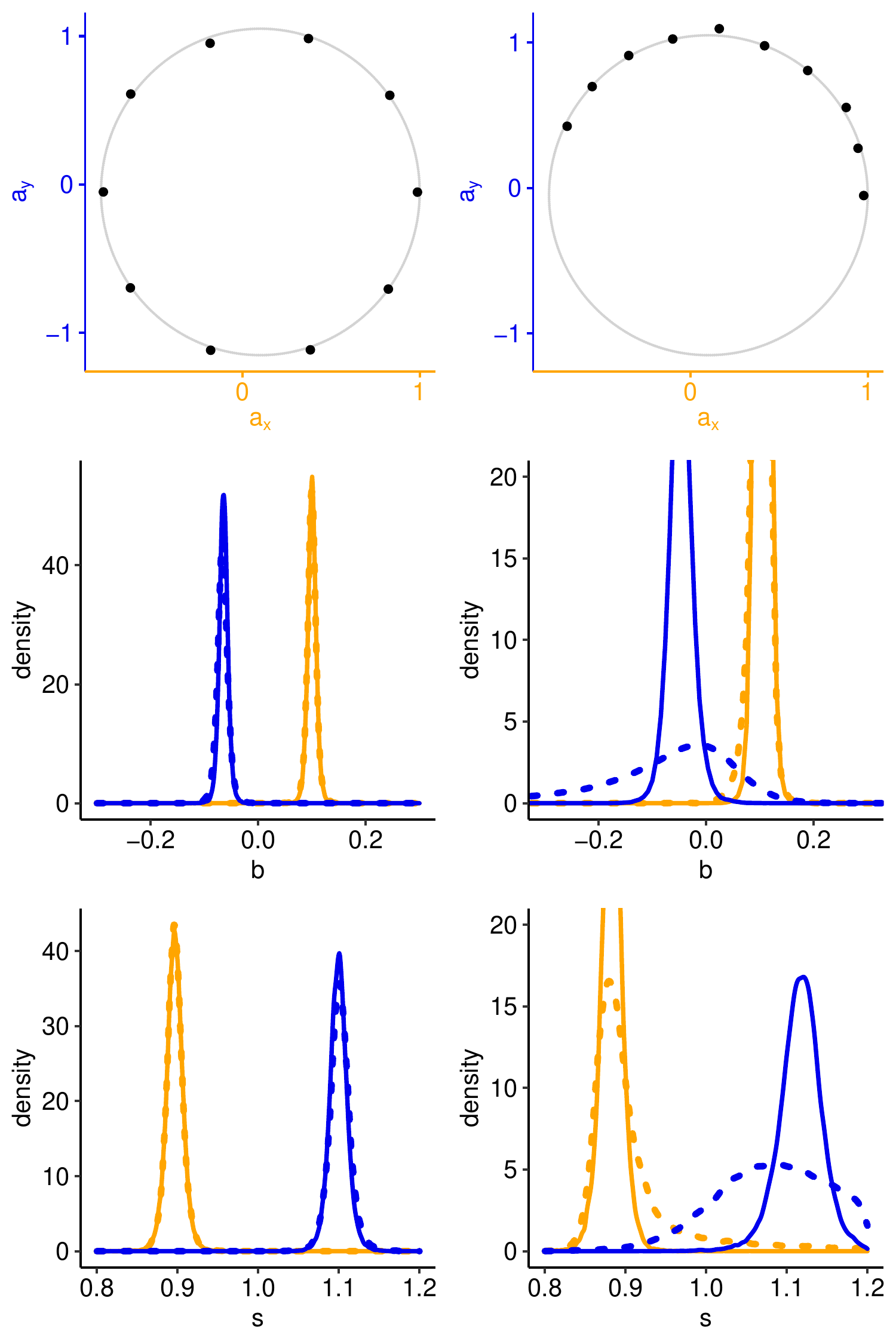}}
\caption{Comparison of the full model (solid line) to the ODR (dashed) approximation by simulation in two dimensions. The upper panel shows the data sampled for the two configurations, both with $\vb=(0.1,-0.05)$ and $\mS=\diag(1.1,0.9)$. The middle and lower panels depict the MCMC density estimates of $\vb$ and $\mS$, respectively. Orange lines represent the $x$ components, and blue lines represent the $y$ components.}
\label{fig:twodim}
\end{figure}

To illustrate the differences between the two methods and demonstrate the capability of Bayesian methods to estimate uncertainty, we performed a 2-dimensional simulation study. We gathered $N=10$ data points from equation \eqref{eq:error_noise2}, using parameters $\vb=(0.1, -0.05)$, $\mS=\diag(0.9,1.1)$, and $\sigma=0.02$. The collected data is displayed in the upper portion of Figure \ref{fig:twodim}. The angles $\varphi$ were distributed evenly throughout the full circumference (left side) or limited (right side).
The marginal densities estimated from the MCMC samples are displayed in the middle and lower portions of \fref{fig:twodim}. When data is collected for the entire circle, the full model and the approximation provide nearly identical results, as seen on the left side of the figure, where the dotted and full lines are difficult to distinguish. However, when data is not collected for the entire circle, as seen on the right side of the figure, the models have less information, and their uncertainties increase, which is reflected in broader distributions. This is particularly evident in the y-component of the parameters, shown in blue and where the data only covers about half of the maximum range. 
As a result, the approximation, which uses one-dimensional radius data instead of full 2D data, has less information and is, therefore, less confident than the full model. It is important to note that all models cover the true values of the parameters, and they can quantify their uncertainty. Bayesian approaches, such as the one used here, inherently provide bounds to their predictions, while classical non-Bayesian methods do not typically provide this information. Further, these results support the intuition that collecting data that spans the maximum possible range is beneficial.
 %%%%%% 2D Comparison %%%%%% 
%%%
% Figure is created with 
% calibration_2D_FullvsODR.qmd
%%%%%%%%%%%%%%%%%%%%%%%%%%%%%%%%%%%%%
\subsection{3D Simulations Results}\label{sec:3D}

In Section \ref{sec:2D}, we illustrated the properties of our Bayesian approaches. Now we want to evaluate the applicability of our Bayesian methods for sensor data. We generated 3D data from a sensor, using specific values for $\vb=(0.1,-0.2,0.3)$, $\mS=\diag(0.9, 1, 1.1)$, and noise $\sigma = 0.02$. We sampled various data points, namely $N=1^2,2^2,3^2,\ldots 20^2$. To mimic a typical recording procedure, we use angles with fixed differences instead of a random sample of angles. Specifically, we use $\sqrt{N}$ evenly spaced angles between $0.1, \ldots 2 \cdot \pi-0.1$ for $\varphi$ and $0.1, \ldots \pi-0.1$ for $\vartheta$. We ran 10,000 samples for a warm-up and used 2,000 samples for estimation. We performed MCMC for the full model until $N=25$ and observed that the credibility intervals for the full model were tighter than the approximation. However, as we reached $N=20$, fitting the full model became challenging. We began to see $\hat{R}$ values larger than $1.2$ and needed to run longer chains. Fortunately, the credibility intervals of the ODR approximation become comparable to the full model around $N=20$. Therefore, we will limit our further discussion to the ODR approximation.

\begin{figure}[!t]
\centerline{\includegraphics[width=.9\columnwidth]{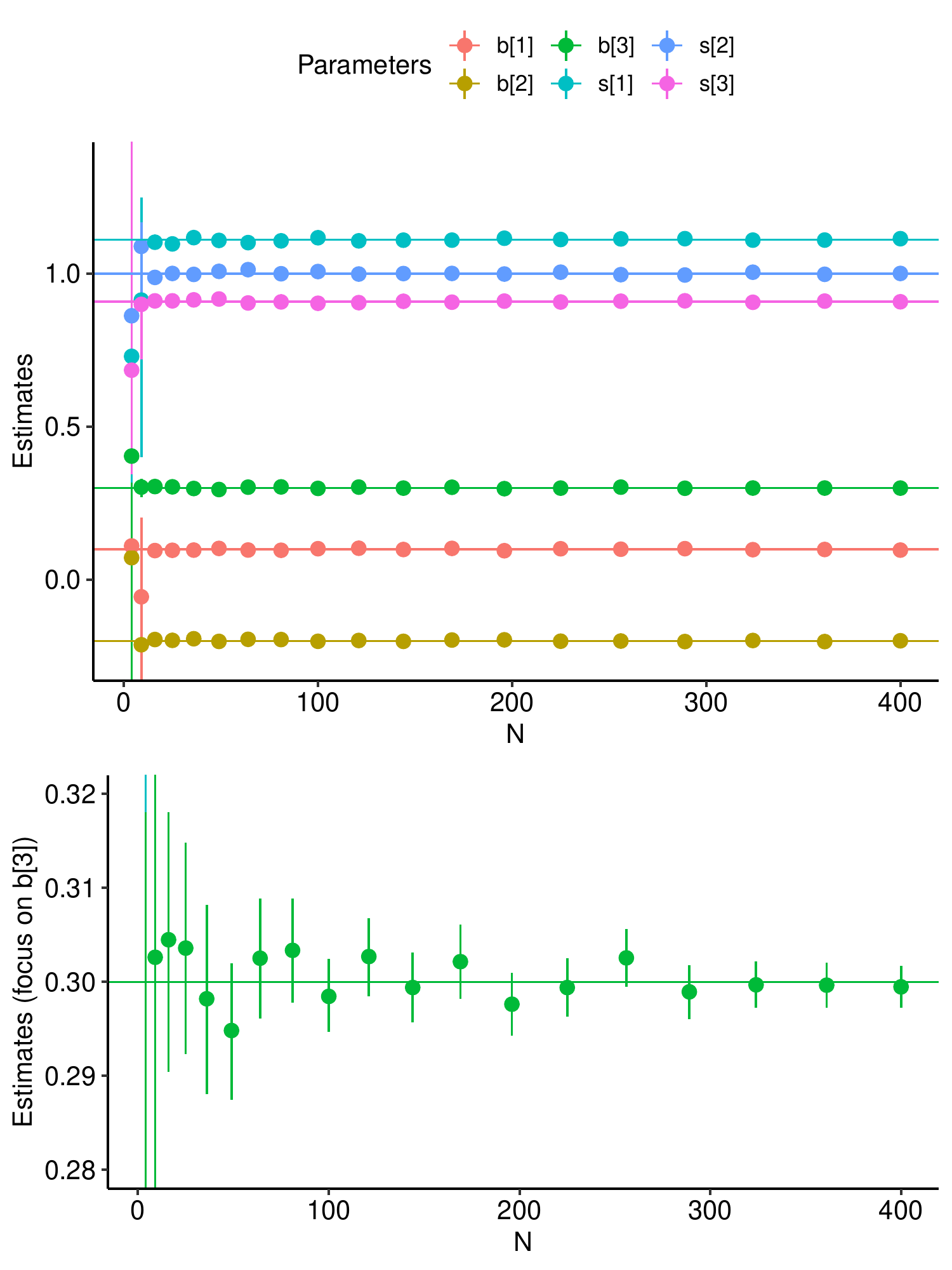}}
\caption{A simulation study for typical sensor data conducted to validate the ODR approximation: The figure shows MCMC estimates of the 90$\%$ credibility intervals (CI), created from the $0.05$ and $0.95$ quantiles and the median (dot) of the parameters $b_j$ and $s_j$ plotted against the number of data points ($N$). The horizontal lines depict the parameters used in the simulation. The lower panel specifically examines the range of the $b_3$ parameter and shows that the deviation from the true values decreases as $N$ increases.}
\label{fig:simu3d}
\end{figure}

For all simulations, we had no difficulty with the convergence of the MCMC samples. Also, the duration of the sampling process is relatively fast, with a maximum of approximately 9 seconds per sample for a sample size of $N=400$ when using four parallel chains on a 2018 MacBook Pro. While this is slower than the maximum likelihood optimization method, which typically takes around 0.1 seconds, it is still sufficiently fast for practical use. However, in contrast to the maximum likelihood approach, we can quantify the uncertainty of the estimated parameters. This is done 
in \fref{fig:simu3d}, showing the $0.05, 0.5$, and $0.95$ quantiles. As expected, the uncertainty decreases with increasing $N$. This is consistent with the expectation that the true values fall within the upper and lower quantiles in approximately 90 percent of the cases, thus forming a valid credibility interval (CI). Furthermore, the median is not systematically above or below the horizontal line, indicating the unbiasedness of our approach. 
%%%%%%% 3D Comparison %%%%%% 
%%%
% Figure is created with 
% 3D_Simu.r ()
%%%%%%%%%%%%%%%%%%%%%%%%%%%%%%%%%%%%%

\subsection{A Real Data Set}\label{sec:real}
\begin{figure}[!t]
\centerline{\includegraphics[width=0.9\columnwidth]{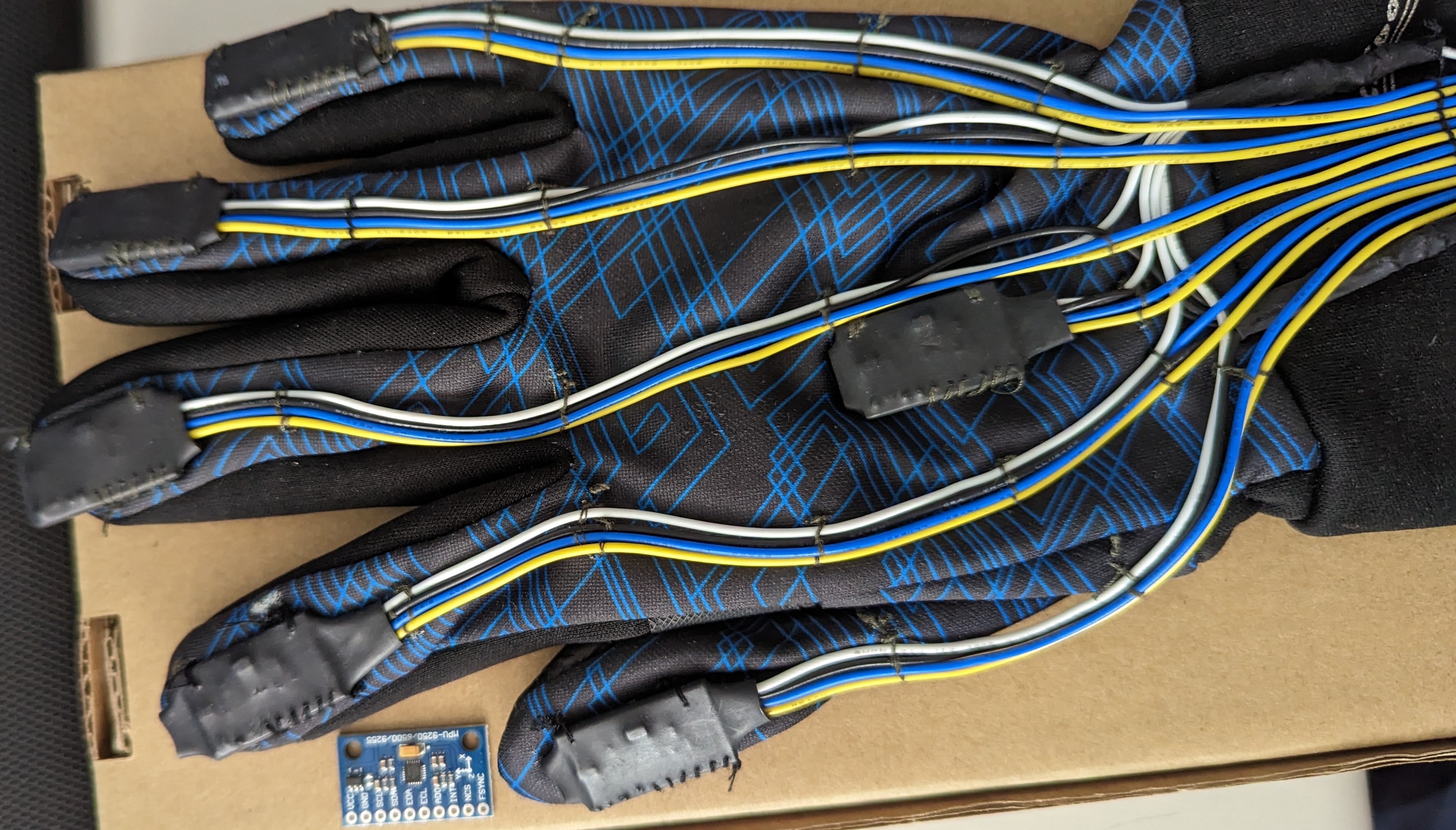}}
\caption{The glove used to demonstrate the calibration of the accelerometer data: Five IMUs are placed on the fingers, approximately at the position of the fingernails, and one IMU is located on the palm. In addition, a single IMU unit is displayed near the thumb for reference.}
\label{fig:gloves}
\end{figure}

\begin{figure}%
\centerline{\includegraphics[width=\columnwidth]{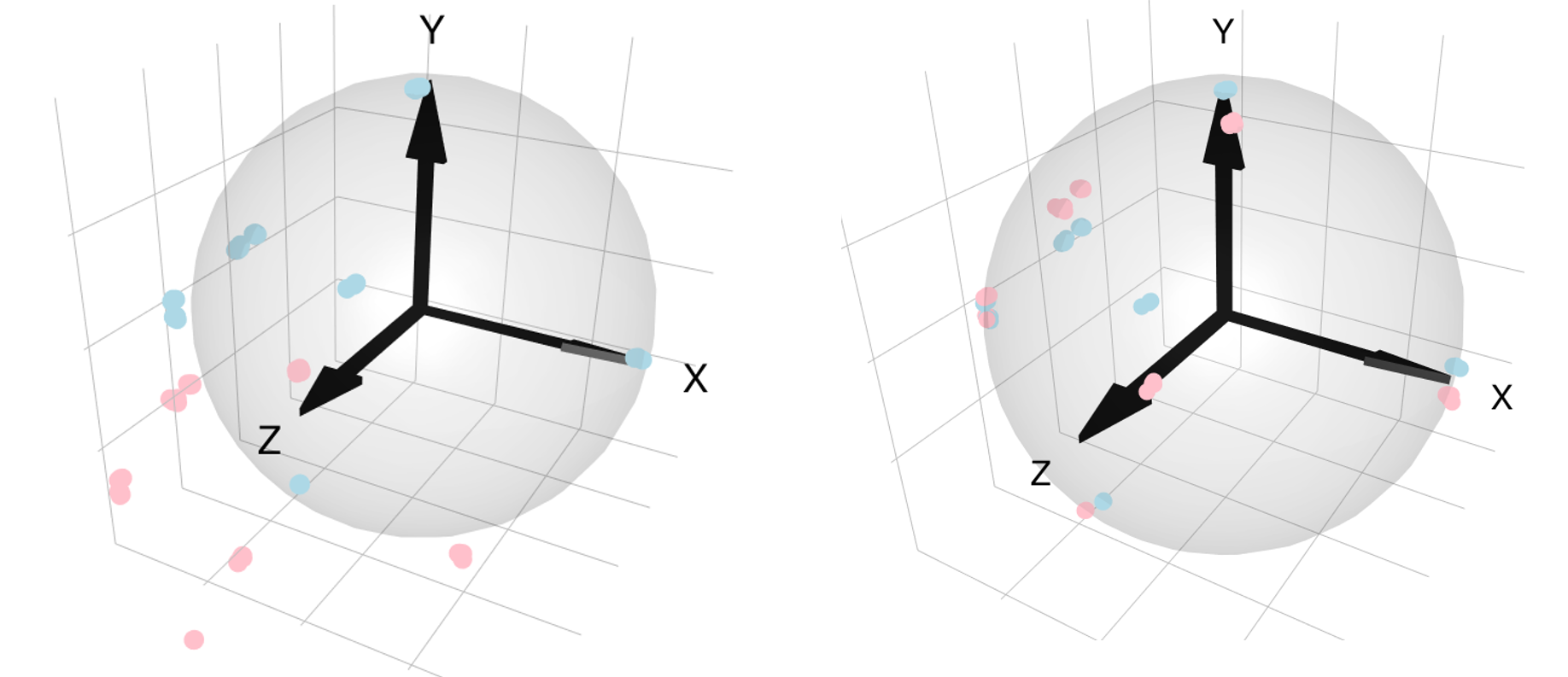}}
\caption{Acceleration data for the IMUs 5 (pink) and 2 (turquoise). The 44 raw data points are shown on the left side. After calibration (right side), the data lies closely on the sphere with radius one (shown in gray).}\label{fig:gloves-sphere} 
\end{figure}

\begin{figure}[!t]
\centerline{\includegraphics[width=\columnwidth]{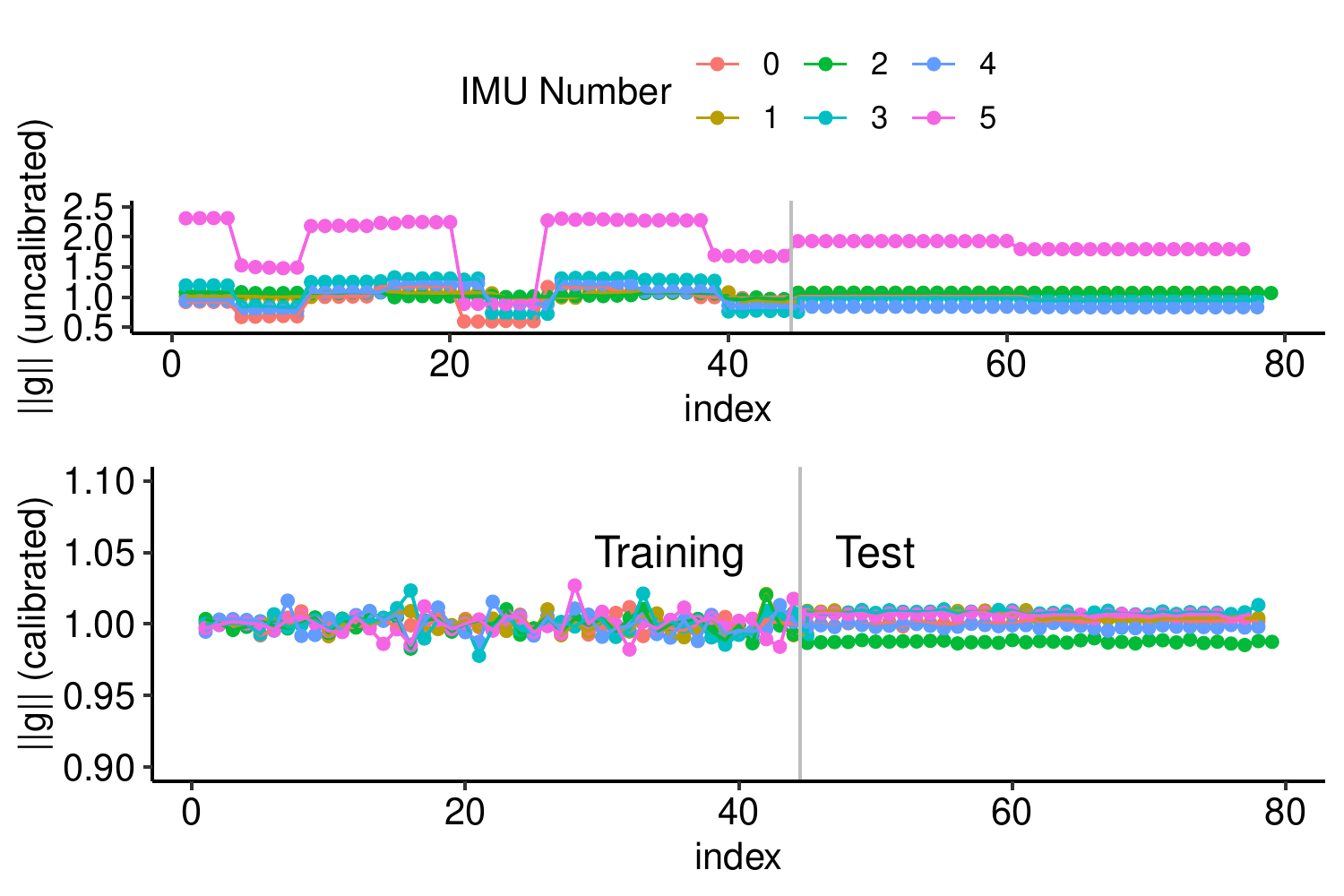}}
\caption{The upper portion of the figure displays the uncalibrated $\sqrt{a_{i1}^2 + a_{i2}^2 + a_{i3}^2}$, and the lower portion shows the calibrated data. The data points used for calibration are shown on the left side of the figure ("training"). The effectiveness of the calibration method is also demonstrated by comparing the calibrated data to data points not used for calibration ("test"), shown on the right side of the figure after the dash.}
\label{fig:gloves-cal}
\end{figure}

The proposed calibration method is demonstrated in a real-world scenario by calibrating the data from the accelerometers of six MPU-9250 IMUs placed on a glove for pose recognition in multimedia applications, as illustrated in \fref{fig:gloves}. The data were collected at a frequency of approximately 5 Hertz using an Arduino and Multiplexer.
%%%%%%% Glove %%%%%% 
%%%
To showcase the robustness of our method in a challenging setting, we recorded the data for calibration by wearing the glove and positioning the hand in various positions, trying to remain motionless without relying on external stabilization devices. The data were recorded for approximately one second, corresponding to approximately 5 data points. The process was repeated by repositioning the glove. The complete data collection took less than two minutes. 
%A video recording of the procedure can be found at \todo{shall we provide the youtube link?}. 
The raw accelerator data for IMU 2 and 5 are shown on the right side of Figure \ref{fig:gloves-sphere}.
%%%%%%% Glove %%%%%%
%%%
% Figure is created from 
% calibration_3D_Glove.qmd and Powerpoint
%%%%%%%%%%%%%%%%%%%%%%%%%%%%%%%%%%%%%
To more clearly visualize the impact of calibration, the upper portion of \fref{fig:gloves-cal} displays $\sqrt{a_{i1}^2 + a_{i2}^2 + a_{i3}^2}$, which should be 1 for calibrated data. The data from the different IMUs exhibit varying degrees of deviation from 1.
%%%%%%% 2D Comparison %%%%%% 
%%%
% Figure is created with 
% calibration_3D_Glove.qmd
%%%%%%%%%%%%%%%%%%%%%%%%%%%%%%%%%%%%%
We experimented with the full model but sometimes encountered difficulties in the convergence of the Markov Chain. Since the ODR approximation converges fast and reliably (the maximum $\hat{R}$ over all parameters and IMUs is 1.0073) and already has a small CI, we used the approximation. In Table \ref{tab:gloves}, the median of the estimated calibration parameters are shown together with the 90$\%$ credibility interval for the different IMUs. The spread of the parameters is roughly 1 percent. If higher precision is needed, more calibration points should be recorded. Further, the glove should not be moved but kept stable. The calibrated accelerations $\va^{\cal}_i$ are obtained from the uncalibrated observations $\va_i$ by using the median of $\mS_{\text{med}}^{-1}$ and the median of the bias $\vb_{\text{med}}$ of the posterior samples via:
\begin{equation}
    \va^{cal}_i = \mS_{\text{med}}^{-1} (\va_i - \vb_{\text{med}}).
\end{equation}
The effect of this calibration using the values from Table \ref{tab:gloves} is shown on the right of \fref{fig:gloves-sphere} and in the lower part of \fref{fig:gloves-cal}. In addition, the effect of the calibration on 
 a novel data set not used for calibration and sampled with stabilization is shown on the right side ("Test"). 
Calibration has a noticeable positive impact, particularly for IMU 5.
%%%%%%% Glove %%%%%% 
\setlength{\tabcolsep}{3pt}
\begin{table}[!h]
\caption{Estimated parameters for the different IMUs}
\label{tab:gloves}
\centering
\begin{tabular}{|r|r|r|r|}
\hline
IMU & $b_1$ & $b_2$ & $b_3$\\
\hline
0 & -0.15 (-0.16, -0.15) & -0.41 (-0.43, -0.39) & -0.14 (-0.14, -0.13)\\
\hline
1 & -0.02 (-0.03, -0.02) & 0.03 (0.01, 0.06) & 0.07 (0.06, 0.07)\\
\hline
2 & -0.03 (-0.03, -0.02) & 0 (-0.01, 0.01) & 0.08 (0.07, 0.09)\\
\hline
3 & -0.31 (-0.31, -0.3) & -0.26 (-0.28, -0.24) & -0.17 (-0.17, -0.16)\\
\hline
4 & -0.09 (-0.09, -0.09) & -0.11 (-0.12, -0.1) & -0.23 (-0.24, -0.22)\\
\hline
5 & -0.88 (-0.88, -0.87) & -1.5 (-1.51, -1.49) & -0.34 (-0.35, -0.32)\\
\hline
IMU & $s_1$ & $s_2$ & $s_3$\\
\hline
0 & 1 (1, 1) & 0.98 (0.96, 1.01) & 0.99 (0.98, 1)\\
\hline
1 & 1 (1, 1) & 1.03 (1.01, 1.06) & 0.99 (0.99, 1)\\
\hline
2 & 1 (0.99, 1) & 1 (0.98, 1.01) & 0.99 (0.98, 1.01)\\
\hline
3 & 1 (1, 1.01) & 1 (0.97, 1.02) & 1 (0.99, 1)\\
\hline
4 & 1 (1, 1) & 1.01 (1, 1.02) & 1 (0.99, 1)\\
\hline
5 & 1 (1, 1.01) & 1 (0.99, 1.01) & 0.98 (0.97, 1)\\
\hline

\multicolumn{4}{p{251pt}}{Shown are the median and the 5 and 95 $\%$ quantiles in brackets. While for scaling, the effect of calibration is small (1 is included in the confidence band), the offsets $b$ vary significantly between the IMUs.}
\end{tabular}
\end{table}
%\begin{figure}[!t]
%\centerline{\includegraphics[width=\columnwidth]{figures/glove_est.pdf}}
%\caption{The estimated parameters for the different IMUs, shown are the 5 and 95 $\%$ quantiles in blue and the median (small red dot).}
%\label{fig:gloves}
%\end{figure}
%%%
% Figure is created with 
% calibration_3D_Glove.qmd
%%%%%%%%%%%%%%%%%%%%%%%%%%%%%%%%%%%%%

\section{Conclusion}
This study has presented a novel Bayesian method for the autonomous calibration of MEMS accelerometer data. The technique has been validated using both artificial and actual data. This method is simple to utilize, as it only requires taking a few measurements of the device in various stationary positions (ideally spanning a broad range) without determining precise angles. Its Bayesian nature allows it to quantify uncertainty naturally. This is an advantage over existing methods, as it can flag problems during data collection or indicate when enough data have been collected for the required calibration precision. The introduced orthogonal distance regression (ODR) approximation mitigates technical difficulties associated with the full Bayesian model when using more than 20 data points. Overall, the simplicity and quantification of uncertainty make this method a strong candidate for replacing existing autonomous calibration procedures.

The proposed Bayesian approach offers excellent flexibility for future extensions. For example, it allows substituting the Gaussian likelihood with a Student's t-distribution to enhance robustness \cite{gelmanBayesianDataAnalysis2015}. Additionally, it offers an easy means of incorporating additional prior knowledge, providing a powerful tool for calibration. 
\appendices
\section{Stan Code}
To unambiguously define the models, we show the Stan code for the full model and the ODR approximation. 
\subsection{Full model}\label{sec:stan-full}
\begin{verbatim}
data {
  int<lower=0> N;
  vector[N] x;
  vector[N] y;
  vector[N] z;
}
parameters {
  row_vector[3] b;
  row_vector<lower=0>[3] s;
  real<lower=0> sigma;// The noise in 
  vector<lower=0,upper= pi()>[N-1] theta;
  vector<lower=0,upper=2*pi()>[N-1] phi;
}
transformed parameters {
  matrix[N, 3] A;
  //Initial conditions theta_0 = phi_0 = 0
  A[1,1] = b[1] ;
  A[1,2] = b[2] ; 
  A[1,3] = b[3] + s[3] * 1; 
  for(i in 2:N){
    A[i,1] = b[1] + s[1] * 
         sin(theta[i-1])*cos(phi[i-1]);
    A[i,2] = b[2] + s[2] * 
        sin(theta[i-1])*sin(phi[i-1]);
    A[i,3] = b[3] + s[3] * 
        cos(theta[i-1]);
  } 
}
model {
  b ~ normal(0,1);
  sigma ~ normal(0,0.2);
  s ~ lognormal(0, 0.5);
  x ~ normal(A[,1],sigma);
  y ~ normal(A[,2],sigma);
  z ~ normal(A[,3],sigma);
}
\end{verbatim}
\subsection{Radial Approximation}\label{sec:stan-ODR}
Note that for convenience, we directly use $S^{-1}$ with the vector $\tt{sinv}$ in the code below.
\begin{verbatim}
data {
  int<lower=0> N;
  int<lower=0> D; //Dimensionality 
  matrix[N, D] X; 
}

parameters {
  row_vector[D] b; 
  real<lower=0> sigma;
  row_vector<lower=0>[D] sinv; 
}

model {
  real mu;
  sigma ~ normal(0,0.2);
  sinv ~ lognormal(0, 0.5);
  b ~ normal(0,1);
  for (i in 1:N){
    mu = 0;
    for (j in 1:D){
      mu += (sinv[j]*(X[i,j] - b[j]))^2;
    }
    target += normal_lpdf(1 | 
    sqrt(mu), sigma);
  }
}
\end{verbatim}

\section*{Acknowledgment}
\begin{CJK*}{UTF8}{bsmi}
Oliver Dürr expresses his gratitude to the members of STUST for their hospitality, particularly 林巧樺 (Lynn), who went above and beyond in organizing his stay at STUST. He also thanks his university for enabling his research semester, with special recognition to Barbara Staehle, Matthias Herman, and Jan Kaiser for covering his lecture duties. Finally, we appreciate the reviewers' and editor's efforts in providing constructive feedback that has improved our paper considerably.
 \end{CJK*}
\section*{References and Footnotes}

\printbibliography
\begin{IEEEbiography}[{\includegraphics[width=1in,height=1.25in,clip,keepaspectratio]{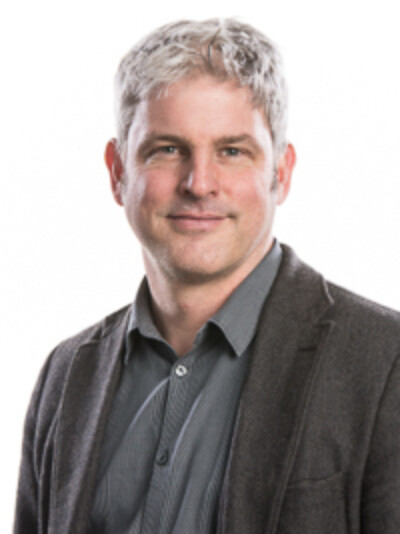}}]
{Oliver Dürr}
  is a professor of data science at the Konstanz University of Applied Sciences. After his Ph.D. in theoretical physics, he worked for ten years in a bioinformatics company, developing and applying machine learning and statistical methods to all kinds of -omics data. He then was a lecturer for statistical data analysis at the Zurich University of Applied Sciences. Currently, he is focused on deep learning, with a particular emphasis on its probabilistic aspect, and is pursuing Applied Bayesian Statistics.
\end{IEEEbiography}

\begin{IEEEbiography}[{\includegraphics[width=1in,height=1.25in,clip,keepaspectratio]{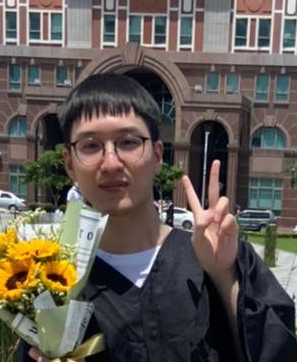}}]{Po-Yu Fan} is a research assistant pursuing his master's degree at the Institute of Computer Science and Information Engineering, Southern Taiwan University of Science and Technology. His current research direction is the application of wearable devices, computer-human interaction, and numerical analysis. 
\end{IEEEbiography}

\begin{IEEEbiography}[{\includegraphics[width=1in,height=1.25in,clip,keepaspectratio]{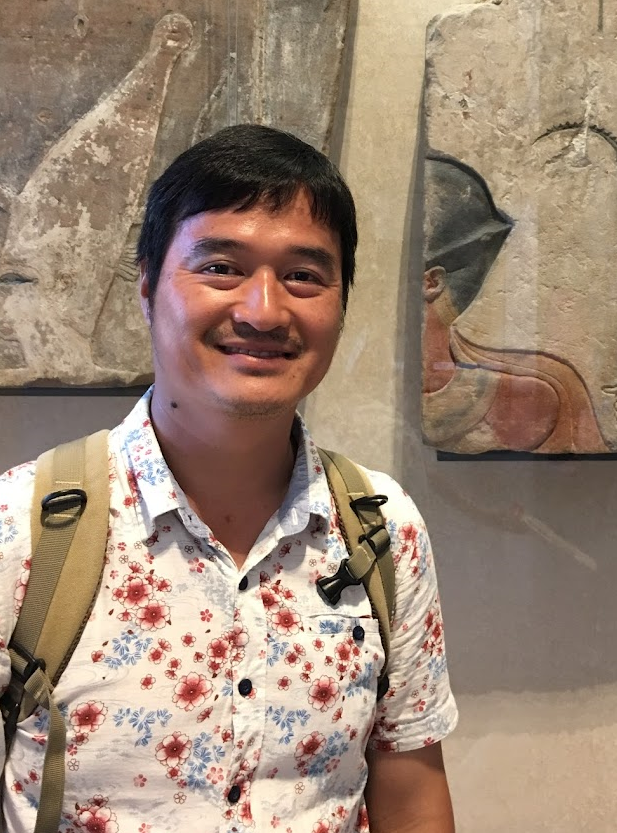}}]{Zong-Xian Yin} is an associate professor in the Department of Computer Science and Information Engineering at the Southern Taiwan University of Science and Technology, Taiwan. He received his Ph.D. in Computer Science and Information Engineering from National Cheng Kung University and his M.S. degree in Industrial Design from the same university. His research interests include computer-human interaction, bioinformatics, machine learning, and virtual and augmented reality.
\end{IEEEbiography}
\end{document}